\documentclass[12pt]{article}
\usepackage{amssymb}

\newcommand{\nc}{\newcommand}
\nc{\la}{\lambda} \nc{\alf}{\alpha}
\nc{\tht}{\theta}  \nc{\be}{\beta}  \nc{\eps}{\epsilon}
\nc{\ga}{\gamma}  \nc{\De}{\Delta}  \nc{\Ga}{\Gamma}  \nc{\vphi}{\varphi}
\nc{\de}{\delta} \nc{\si}{\sigma}  \nc{\ka}{\kappa}  \nc{\T}{\Theta}
\nc{\om}{\omega}  \nc{\Om}{\Omega}  
\nc{\qq}{\quad\quad}  \nc{\nf}{\infty}   \nc{\dl}{\mathop{\smash{\cal L}}}
\nc{\ra}{\rightarrow}  \nc{\ol}{\overline}  \nc{\und}{\underline}
\nc{\beq}{\begin{equation}}   \nc{\eeq}{\end{equation}}
\nc{\beqa}{\begin{eqnarray}}  \nc{\eeqa}{\end{eqnarray}}
\nc{\nin}{\noindent}          \nc{\pt}{\partial}
\nc{\dst}{\displaystyle}      \nc{\nnb}{\nonumber}
\nc{\bs}{\backslash}          \nc{\mb}{\mathbb} 
\nc{\dg}{\dagger}   \nc{\wh}{\widehat}  \nc{\wti}{\widetilde}
\nc{\PT}{\Pi_{\tht}}  \nc{\PF}{\Pi_{\phi}}       \nc{\R}{\mb R}

\newcounter{muni}
\newenvironment{remunerate}{\begin{list}{{\rm \arabic{muni}.}}
{\usecounter{muni}
\setlength{\leftmargin}{0pt}\setlength{\itemindent}{38pt}}}{\end{list}}

\nc{\brm}{\begin{remunerate}}   \nc{\erm}{\end{remunerate}}
\nc{\barr}{\begin{array}}   \nc{\earr}{\end{array}}

\nc{\stg}{\mathop{\smash{*}}}    \nc{\st}{\mathop{\smash{\delta}}}
 
\nc{\mtvb}{\mathversion{bold}}   \nc{\mtvn}{\mathversion{normal}}

\begin{document}

\begin{titlepage}

\[  \]
\centerline{\Large\bf Bianchi III and V Einstein metrics}

\vskip 2.0truecm
\centerline{\large\bf Galliano VALENT${}^{\;\dagger\; *}$}

\vskip 2.0truecm
\centerline{${}^{\dagger}$ \it Laboratoire de Physique Th\'eorique et des
Hautes Energies}
\centerline{\it CNRS, Unit\'e associ\'ee URA 280}
\centerline{\it 2 Place Jussieu, F-75251 Paris Cedex 05, France}
\nopagebreak

\vskip 0.5truecm
\centerline{${}^*$ \it D\'epartement de Math\'ematiques}
\centerline{\it UFR Sciences-Luminy}
\centerline{\it Case 901 163 Avenue de Luminy}
\centerline{\it 13258 Marseille Cedex 9, France}
\nopagebreak

\vskip 2.5truecm

\begin{abstract}
We present diagonal Einstein metrics for Bianchi III and V, both for minkowskian and 
euclidean signatures and we show that the Einstein Bianchi III metrics have an integrable 
geodesic flow. 
\end{abstract}

\end{titlepage}

\section{Introduction}
As is clear from the book by Stephani et al \cite{sk}, the field of exact solutions to 
Einstein's field equations has been enriched substantially, and quite recently impressive 
progresses have taken place for ricci-flat metrics within the Bianchi B family. In 
\cite{ct} solutions were obtained for some non-diagonal Bianchi III metrics, and  
in \cite{tc} for the more general case of Bianchi VII$_h$. 

The aim of this article is to obtain some new exact Einstein metrics, for both minkowskian 
and euclidean signatures. These metrics are obtained for 
the simplest Bianchi B metrics: the type III and the type V, under the simplifying 
assumption that they are diagonal with respect to the invariant 1-forms.

The content of this article is the following: in Section 2 we present background informations 
and the field equations for the Bianchi III metrics. In Section 3 we derive the Einstein metric 
and prove the integrability  of the geodesic flow; quite remarkably this metric exhibits  
a fourth Killing vector. Then section 4 presents background informations and the field equations 
of the Bianchi V diagonal metrics. They are first solved, in Section 5, for the ricci-flat case to 
recover Joseph's metric and its euclidean partner. The basic integration of the Einstein 
equations is given in Section 6. Due to the higher complexity of these metrics, which do 
involve elliptic functions, the explicit forms are given in Section 7 for the minkowskian 
signature and in Section 8 for the euclidean signature. Some conclusions are presented in 
Section 9. We give in the appendices more details on a curious form of de Sitter metric 
encountered in the analysis of the Bianchi V Einstein metrics and some technicalities related 
to elliptic functions.

\section{Bianchi III metrics}
The Bianchi III Lie algebra is defined as
\beq\label{3lie}
[{\cal L}_1,{\cal L}_2]=0,\qq [{\cal L}_2,{\cal L}_3]=0,\qq [{\cal L}_3,{\cal L}_1]={\cal L}_3.\eeq
A representation by differential operators is
\beq\label{3iso}
{\cal L}_1=\pt_x+z\,\pt_z,\qq {\cal L}_2=\pt_y,\qq {\cal L}_3=\pt_z,\eeq
and the invariant Maurer-Cartan 1-forms are
\beq\label{3mc}
\si_1=dx,\quad\si_2=dy,\quad \si_3=e^{-x}\,dz,\quad\Longrightarrow\quad  
d\si_1=d\si_2=0,\quad d\si_3=\si_3\wedge\si_1.\eeq
We will look for diagonal metrics of the form
\beq\label{metb3}
g=\beta^2\,\si_1^2+\ga^2\,\si_2^2+\de^2\,\si_3^2+\eps\,\alpha^2\,dt^2.
\eeq
\subsection{Flat space}
Before writing down the Einstein equations, it is interesting to look for flat space within our 
cooordinates choice. An easy computation shows that it is given by
\beq\label{plateb3}
g_0=\si_2^2+t^2(\si_1^2+\si_3^2)-dt^2=dy^2+t^2\,\frac{(dz^2+dr^2)}{r^2}-dt^2,\qq r=e^x.
\eeq
This metric is {\em unique} and does exist only for the minkowskian signature. The flattening coordinates are
\[x_1=y,\qq x_2=\frac{tz}{r},\qq x_3=\frac t{2r}(-1+z^2+r^2),\qq \tau=\frac t{2r}(1+z^2+r^2),\]
which gives
\[g_0=d\vec{r}\cdot d\vec{r}-d\tau^2,\qq\quad \vec{r}=(x_1,x_2,x_3).\]
The appearance of $\,\si_1^2+\si_3^2$ signals an extra symmetry  
\beq\label{iso4}
{\cal L}_4=z\,\pt_x+\frac 12\,(z^2-e^{2x})\pt_z,\eeq
which enlarges the infinitesimal isometries to the four dimensional Lie algebra\footnote{We give 
only the non-vanishing commutators.}
\beq
[{\cal L}_3,{\cal L}_1]={\cal L}_3,\qq [{\cal L}_1,{\cal L}_4]={\cal L}_4,\qq  
[{\cal L}_3,{\cal L}_4]={\cal L}_1.\eeq

\subsection{The field equations}
The Einstein equations \footnote{In our notations the spheres have positive curvature.}
\[Ric_{~\mu}^{~\nu}=\la\,\de_{~\mu}^{~\nu}\] 
give for the Bianchi III case 
\[\barr{ll} 
(I)\quad &\dst \frac{\dot{\de}}{\de}=\frac{\dot{\be}}{\be}, \\[5mm]  
(II)\quad & \dst\frac{\ddot{\be}}{\be}+\frac{\dot{\be}}{\be}\left(\frac{\dot{\be}}{\be}
+\frac{\dot{\ga}}{\ga}-\frac{\dot{\alf}}{\alf}\right)+\eps\left(\frac 1{\be^2}+\la\right)\alf^2=0,\\[5mm]
(III)\quad & \dst \frac{\ddot{\ga}}{\ga}+\frac{\dot{\ga}}{\ga}\left(2\frac{\dot{\be}}{\be}
-\frac{\dot{\alf}}{\alf}\right)+\eps\,\la\,\alf^2=0, \\[5mm]    
(IV)\quad & \dst \frac{\dot{\be}^2}{\be^2}+2\frac{\dot{\be}\dot{\ga}}{\be\ga}
+\eps\left(\frac 1{\be^2}+\la\right)\alf^2=0.\earr\]

\section{Einstein metrics and their geodesic flow}
Let us now consider the $\la\neq 0$ case. Relations (I) and (II)-(IV) integrate up to
\beq\label{int1}
\frac{\dot{\be}}{\be}=c\,\frac{\alf\,\ga}{\be},\qq c\in{\mb R},\qq\qq \de=c_2\,\be,\qq c_2\neq 0,\qq\qq .\eeq
The coordinate choice 
\[\alf=\frac{\be}{\ga}\quad\Longrightarrow\quad \be=\be_0\,e^{c t},\qq \de=c_2\,\be_0\,e^{c t}.\]
To determine $\ga$ we have to use (III) which becomes 
\beq\label{int2}
\frac{\ddot{\ga}}{\ga}+\frac{\dot{\ga}^2}{\ga^2}+c\,\frac{\dot{\ga}}{\ga}+
\eps\la\be_0^2\,\frac{e^{2c t}}{\ga^2}=0.
\eeq
This equation does linearize in $\ga^2$ to
\beq\label{int3}
\ddot{(\ga^2)}+c\,\dot{(\ga^2)}+2\eps\la\be_0^2\,e^{2c t}=0,
\eeq
and the remaining relation (IV) becomes
\beq\label{int4}
c\,\dot{(\ga^2)}+c^2\,\ga^2+\eps(1+\la\,\be^2)=0.
\eeq
Let us organize the discussion according to the values of $c$.

\subsection{Vanishing $c$}
The relation (\ref{int4}) gives 
$\,\be^2=-1/\la$ and (\ref{int3}) is easily integrated to $\ga^2=\ga_0+\ga_1 t+\eps t^2$. 
By a translation of $t$ we can set $\ga_1\to 0$ and by a rescaling of $z$ we can set $c_2\to 1$, 
so we can write the metric
\beq\label{MEfin1} 
g=\frac 1{|\la|}\Big[\si_1^2+\si_3^2+\ga^2\,\si_2^2+\eps\frac{dt^2}{\ga^2}\Big],\qq 
\ga^2=\ga_0+\eps t^2,\qq \la<0.
\eeq
Let us notice that all the metrics will have $\la<0$.

For the minkowskian signature we must have $\ga_0>0$ and the change of variable $\,t=\sqrt{\ga_0}\,{\rm th}\,\tau$ tansforms the metric into
\beq
g=\frac 1{|\la|}\left\{\si_1^2+\si_3^2+\frac 1{{\rm ch}^2\,\tau}\Big[\si_2^2-d\tau^2\Big]\right\}.
\eeq
We get a decomposable space-time \cite{sk}[p. 554] which is the product of two 2-dimensional Einstein metrics: ${\mb H}^2$ on the one hand with metric and isometries
\beq
g_0=\si_1^2+\si_3^2=dx^2+e^{-2x}\,dz^2,\qq \left\{\barr{l}L_1=\pt_z,\qq L_2=\pt_x+z\pt_z,\\[4mm]  
L_3=z\pt_x+\frac 12(z^2-e^{2x})\pt_z,\earr\right.
\eeq
and a Lorentzian 2-dimensional metric on the other hand with isometries
\beq
g_1=\frac{dy^2-d\tau^2}{{\rm ch}^2\,\tau},\qq \left\{\barr{l} M_1=\pt_y,\qq M_2=e^{-y}({\rm sh}\,\tau\,\pt_y-{\rm ch}\,\tau\,\pt_{\tau}),\\[4mm] M_3=e^{y}({\rm sh}\,\tau\,\pt_y+{\rm ch}\,\tau\,\pt_{\tau}).\earr\right.
\eeq
In this very special case we have as many as 6 Killing vectors!

For the euclidean signature, according to the sign of $\ga_0$ we have 3 cases:
\[\barr{ll}
\ga_0>0\qq & \dst g=\frac 1{|\la|}\left\{\si_1^2+\si_3^2
+\frac 1{\cos^2 \tau}\Big[\si_2^2+d\tau^2\Big]\right\},\\[5mm]
\ga_0=0\qq & \dst g=\frac 1{|\la|}\left\{\si_1^2+\si_3^2+\frac 1{\tau^2}\Big[\si_2^2+d\tau^2\Big]\right\},\\[5mm] 
\ga_0<0\qq & \dst  g=\frac 1{|\la|}\left\{\si_1^2+\si_3^2
+\frac 1{{\rm sh}^2\,\tau}\Big[\si_2^2+d\tau^2\Big]\right\},
\earr\qq\quad \la<0.\]
We have again decomposable Einstein metrics made up of two copies of ${\mb H}^2$.

\subsection{Non-vanishing $c$}
In this more general case we obtain  
\beq
\Ga^2\equiv c_2^2\,\ga^2=-\eps+\ga_1\,e^{-c t}-\frac{\eps\la\be_0^2}{3}\,e^{2c t}.
\eeq
Taking as variable $s=\be_0\,e^{c t}$, and cleaning up the irrelevant parameters, 
we eventually obtain the Einstein metric
\beq\label{MEfin2}
g=s^2(\si_1^2+\si_3^2)+\Ga^2\,\si_2^2+\eps\,\frac{ds^2}{\Ga^2},\qq
\Ga^2=-\eps+\frac{\ga_0}{s}-\frac{\eps\la}{3}\,s^2.
\eeq
This metric exhibits the extra Killing vector ${\cal L}_4$ defined in (\ref{iso4}) but it is 
no longer decomposable.

\subsection{Integrable geodesic flow}
The Einstein metric that we have found is type D. If it were a vacuum metric, the existence of 
one Killing-Yano tensor and of one Killing-St\"ackel tensor \footnote{We follow the same 
terminology as in \cite{vy}.} would follow from \cite{wp}, \cite{Co}. Nevertheless, for 
the obvious tetrad, we found the following Killing-Yano tensor for metrics with both signatures
\beq
Y=s\,e_3\wedge e_1.\eeq
Its square gives the Killing-Stackel tensor
\beq\label{4ks}
S=s^2\Big(e_1^2+e_3^2\Big).\eeq
Let consider, for the minkowskian signature (the euclidean case is similar), the geodesic flow 
induced by the Hamiltonian 
\beq
H=\frac 12\left(\frac 1{\Ga^2}\,\Pi_y^2+\frac{\Pi_x^2+e^{2x}\Pi_z^2}{s^2}
-\Ga^2\,\Pi_s^2\right).
\eeq
The KS tensor (\ref{4ks}) gives for conserved quantity
\beq
Q=\Pi_x^2+e^{2x}\Pi_z^2,\qq\quad \{H,Q\}=0.
\eeq
It cannot be obtained from symmetrized tensor products of Killing vectors because their 
corresponding linear conserved quantities are
\[\wti{\cal L}_1=\Pi_y, \quad \wti{\cal L}_2=\Pi_z,\quad \wti{\cal L}_3=\Pi_x+z\,\Pi_z,\quad 
\wti{\cal L}_4=z\,\Pi_x+\frac 12\Big(z^2-e^{2x}\Big)\Pi_z,\quad \{H,\wti{\cal L}_i\}=0.\]
The dynamical system with hamiltonian $H$ is therefore integrable, since it exhibits 4 
independent conserved quantities: $\,H,\,Q,\,\Pi_y,\,\Pi_z$ in involution for the 
Poisson bracket. Writing the action as
\beq
S=E\,t+p\,\Pi_y+q\,\Pi_z+A(s)
\eeq 
we get for separated equation
\beq
\left(\frac{dA}{ds}\right)^2=\frac{Q}{s^2\Ga^2}+\frac{p^2}{\Ga^4}-\frac{2E}{\Ga^2}.
\eeq
Notice that since these metrics are Einstein, the ``minimal quantization" discussed in 
\cite{dv} does preserve integrability at the quantum level and implies the separability of the 
Schr\"{o}dinger equation. 

\subsection{The ricci-flat limit}
The coordinates used in (\ref{MEfin2}) allow to take the $\la\to 0$ limit, giving
\beq
g_0=s^2(\si_1^2+\si_3^2)+\ga^2\,\si_2^2+\eps\,\frac{ds^2}{\ga^2},\qq \ga^2=-\eps+\frac{\ga_0}{s}.
\eeq
Of course, this metric is certainly not new since it is type D: it must lie somewhere in Kinnersley analysis \cite{Ki} of all ricci-flat minkowskian type D metrics. Obviously its geodesic flow 
is also integrable.

\section{Bianchi V}
In this case the Lie algebra is
\beq\label{5lie}
[{\cal L}_1,{\cal L}_2]={\cal L}_2,\qq [{\cal L}_2,{\cal L}_3]=0,\qq [{\cal L}_3,{\cal L}_1]=-{\cal L}_3,\eeq
with the Killing vectors 
\beq\label{5iso}
{\cal L}_1=\pt_x-y\pt_y-z\pt_z,\qq {\cal L}_2=\pt_y,\qq {\cal L}_3=\pt_z,\eeq
and the invariant Maurer-Cartan 1-forms  
\beq\label{5mc}
\si_1=dx,\ \si_2=e^x\,dy,\  \si_3=e^x\,dz,\quad\Rightarrow\quad 
d\si_1=0,\quad d\si_2=\si_1\wedge\si_2,\quad d\si_3=\si_1\wedge\si_3.\eeq
We will look again for a diagonal metric
\beq\label{5met}
g=\be^2\,\si_1^2+\ga^2\,\si_2^2+\de^2\,\si_3^2+\eps\,\alf^2\,dt^2.\eeq
\subsection{The flat space}
Let us first determine the flat space Bianchi V metric. It is easy to check that it is given by
\beq\label{b5plate}
g_0=t^2(\si_1^2+\si_2^2+\si_3^2)-dt^2=t^2\,\ga-dt^2.
\eeq
This metrique is {\em unique} and does exist only with the minkowskian signature. The metric $\ga$ is 
easily seen to be the Poincar\'e metric for ${\mb H}^3$, since we can write
\[\ga\equiv \si_1^2+\si_2^2+\si_3^2=\frac{dy^2+dz^2+d\rho^2}{\rho^2},\qq \rho=e^{-x},\]
which has 6 Killing vectors. The flattening coordinates 
for (\ref{b5plate}) are
\beq\label{b5plate1}
x_1=\frac{ty}{\rho},\quad x_2=\frac{tz}{\rho},\quad x_3=\frac t{2\rho}(-1+y^2+z^2+\rho^2),
\quad \tau=\frac t{2\rho}(1+y^2+z^2+\rho^2),\eeq
leading to 
\beq\label{b5plate2}
g_0\equiv t^2(\si_1^2+\si_2^2+\si_3^2)-dt^2=d\vec{r}\cdot d\vec{r}-d\tau^2,
\qq \vec{r}=(x_1,x_2,x_3).\eeq

\subsection{The field equations}
The Einstein equations for Bianchi V are
\beq\label{b5E}\barr{ll}
(I)\quad & \dst\frac{\ddot{\be}}{\be}+\frac{\dot{\be}}{\be}
\left(\frac{\dot{\ga}}{\ga}+\frac{\dot{\de}}{\de}-\frac{\dot{\alf}}{\alf}\right)
+\eps(2+\la\,\be^2)\frac{\alf^2}{\be^2}=0,\\[5mm]
(II)\quad & \dst\frac{\ddot{\ga}}{\ga}+\frac{\dot{\ga}}{\ga}
\left(\frac{\dot{\be}}{\be}+\frac{\dot{\de}}{\de}-\frac{\dot{\alf}}{\alf}\right)
+\eps(2+\la\,\be^2)\frac{\alf^2}{\be^2}=0,\\[5mm]
(III)\quad & \dst\frac{\ddot{\de}}{\de}+\frac{\dot{\de}}{\de}
\left(\frac{\dot{\be}}{\be}+\frac{\dot{\ga}}{\ga}-\frac{\dot{\alf}}{\alf}\right)
+\eps(2+\la\,\be^2)\frac{\alf^2}{\be^2}=0,\\[5mm]
(IV)\quad & \dst \frac{\dot{\be}\dot{\ga}}{\be\ga}+\frac{\dot{\ga}\dot{\de}}{\ga\de}+
\frac{\dot{\be}\dot{\de}}{\be\de}+\eps(3+\la\,\be^2)\frac{\alf^2}{\be^2}=0,\earr  \quad (V)\quad 
\frac{\dot{\de}}{\de}-2\frac{\dot{\be}}{\be}+\frac{\dot{\ga}}{\ga}=0.
\eeq
We will begin by the Ricci-flat case.

\section{Ricci-flat metrics}
The most general metric, due to Joseph \cite{Jo}, is well known, but as a warming up, let us present 
a new short derivation. Let us put $\la=0$ in (\ref{b5E}); the differences (I)-(II) and (III)-(I) integrate to
\beq
\frac{\dot{\be}}{\be}-\frac{\dot{\ga}}{\ga}=c\,\frac{\alf}{\be\ga\de},\qq\quad 
\frac{\dot{\de}}{\de}-\frac{\dot{\be}}{\be}=c_2\,\frac{\alf}{\be\ga\de},\eeq
and (V) implies $c_2=-c$.

The coordinates choice  
\[\alf=\be\ga\de\quad\Longrightarrow\quad \ga=\ga_0\,e^{c t}\,\be,\qq \de=\de_0\,e^{-c t}\,\be.\]
Let us notice that $\ga_0$ (resp. $\de_0$) can be absorbed in a re-definition of the coordinate 
$y$ (resp. $z$) appearing in $\si_2$ and $\si_3$, so we will take 
$\ga_0=\de_0=1$ in what follows. This remark allows to write the metric
\[g=\be^2\Big(\si_1^2+e^{2c t}\,\si_2^2
+e^{-2c t}\,\si_3^2+\eps\,\,\be^4\,dt^2\Big).\]
Relation (I) becomes
\beq\label{EDbeta}
D_t\,\left(\frac{\dot{\be}}{\be}\right)+2\eps\,\be^4=0,\qq\Longrightarrow\qq \frac{\dot{\be}^2}{\be^2}+\eps\,\be^4=E.
\eeq
Then relation (IV) gives $E=c^2/3$. 

In the minkowskian case, we may have $E=c=0$. This implies the relation 
$dt^2=d\be/\be^6$, and using $\,\be=s$ as a new variable we recover the 
flat metric (\ref{b5plate}).

For $c_1 \neq 0,$ using as a new variable $\dst u=\frac{\sqrt{3}}{c}\,\be^2$ we get
\beq
\frac{du}{\sqrt{1+u^2}}=\pm \frac{2c}{\sqrt{3}}\,dt\qq\Longrightarrow\qq
u={\rm sh}\,[2c(t-t_0)/\sqrt{3}],
\eeq
and, setting $c=1,\ t_0=0$, we have 
\[g=\frac 1u\left[\si_1^2+(u+\sqrt{1+u^2})^{-\sqrt{3}}\,\si_2^2+
(u+\sqrt{1+u^2})^{\sqrt{3}}\,\si_3^2-\frac{du^2}{4u^2(1+u^2)}\right],\quad u\in(0,+\nf).\]
Switching to the new variable $\tau$ we 
eventually obtain 
\beq\label{b5diagM}
{\rm sh}\,(2\tau)=\frac 1{u}\quad\Rightarrow\quad  g={\rm sh}\,(2\tau)\left[\si_1^2+({\rm th}\,\tau)^{\sqrt{3}}\,\si_2^2+({\rm th}\,\tau)^{-\sqrt{3}}\,\si_3^2-d\tau^2\right],
\eeq
the standard form of the minkowskian Joseph metric. Due to the symmetric role played by 
$\,(\si_2,\,\si_3)$, the coefficients of $\si_2^2$ and of $\si_3^2$ may be interchanged, and 
this corresponds to the exchange $\,(c\leftrightarrow -c)$.

For the euclidean Joseph metric we get merely
\beq\label{b5diagE}
g=\sin(2\tau)\left[\si_1^2+(\tan\tau)^{\sqrt{3}}\,\si_2^2+
(\tan\tau)^{-\sqrt{3}}\,\si_3^2+d\tau^2\right], \qq \tau\in(0,\pi/2),
\eeq
and there is no special case $E=0$. 

\section{Einstein Bianchi V metrics}
Let us consider a non-vanishing $\la$. The differences (I)-(II) and (I)-(III) integrate to
\[\frac{\dot{\ga}}{\ga}-\frac{\dot{\be}}{\be}=c\,\frac{\alf}{\be\ga\de},\qq\quad 
\frac{\dot{\de}}{\de}-\frac{\dot{\be}}{\be}=c_2\,\frac{\alf}{\be\ga\de},\]
and (V) implies $c_2=-c$.

The coordinates choice 
\beq\label{5int1}
\alf=\be\ga\de\quad\Longrightarrow\quad \ga=\ga_0\,e^{c t}\,\be,\qq \de=\de_0\,e^{-c t}\,\be,\qq \alf=\ga_0\,\de_0\,\be^3.
\eeq
By the same argument as for the ricci-flat case we may set $\ga_0=\de_0=1$ 
and relation (I) becomes
\beq\label{5int2}
D_t{\left(\frac{\dot{\be}}{\be}\right)}+\eps\,\be^4(2+\la\,\be^2)=0,
\quad\Longrightarrow\quad \frac{\dot{\be}^2}{\be^2}+\eps\,\be^4(1+\la\,\be^2/3)=E.
\eeq
Eventually relation (IV) gives $E=c^2/3\geq 0$. 

\subsection{The special case $\,E=c=0$}
Relation (\ref{5int2}) becomes
\beq
dt=\frac{d\be}{\be^3\sqrt{-\eps-\eps\la\be^2/3}}.\eeq
Taking $\be\to s$ as a  new variable, we get the metric
\beq\label{Esimple}
g=s^2\Big(\si_1^2+\si_2^2+\si_3^2\Big)-\frac{ds^2}{1+\frac{\dst\la\,s^2}{3}}.
\eeq
The minkowskian or euclidean character of the metric does depend solely on the range taken by the 
variable $t$, and in the $\la\to 0$ limit we recover, as it should, the flat space metric (\ref{b5plate}). 

For $\la>0$, we can have only a minkowskian metric. As already experienced with the 
special $c=0$ case for Bianchi III, we may expect some higher symmetry and it is indeed 
the case! Defining $\dst \sqrt{\frac{\la}{3}}\,s=\frac{2t}{1-t^2}$ 
we can write the metric:
\[g^+_M=\frac{12}{\la}\frac 1{(1-t^2)^2}\Big(t^2(\si_1^2+\si_2^2+\si_3^2)-dt^2\Big),\]
on which we recognize a symmetric space, since by using the flattening coordinates (\ref{b5plate1}), we have
\[g^+_M=\frac{12}{\la}\frac{d\vec{r}\cdot d\vec{r}-d\tau^2}{(1+\vec{r}\,^2-\tau^2)^2}.\]
Indeed, using the constrained coordinates 
\[z_0=\frac{1-\vec{r}\,^2+\tau^2}{1+\vec{r}\,^2-\tau^2},\qq 
\vec{z}=\frac{2\vec{r}}{1+\vec{r}\,^2-\tau^2},
\qq z_4=\frac{2\tau}{1+\vec{r}\,^2-\tau^2},\qq z_0^2+\vec{z}\,^2-z_4^2=1,\]
we see that we end up with de Sitter metric
\[g^+_M=\frac 3{\la}\Big(dz_0^2+d\vec{z}\cdot d\vec{z}-dz_4^2\Big),\]
and the isometry group enlarges to $O(4,1)$. In some sense the metric (\ref{Esimple}) is 
an exotic but simple way of writing de Sitter metric and some further details are gathered in 
Appendix A.

For $\la<0$ we have, for Minkowskian signature, anti de Sitter metric
\[ g^-_M=\frac{12}{|\la|(1+t^2)^2}\,\Big(t^2(\si_1^2+\si_2^2+\si_3^2)-dt^2\Big).\]
and a euclidean one
\[g^-_E=\frac 3{|\la|}\Big[{\rm ch}\,^2\tht(\si_1^2+\si_2^2+\si_3^2)+d\tht^2\Big],\]
which is also a symmetric space but we were not able to put a name on it.

\subsection{The general case $\,E\neq 0$}
In relation (\ref{5int2}), let us introduce as a new variable
\beq\label{deft}
\rho=\frac{|c|}{\be^2}>0\quad\Longrightarrow\quad  
\frac{\rho\,d\rho}{\sqrt{P(\rho)}}=\pm\frac{2c\,dt}{\sqrt{3}},\qq\qq P(\rho)\equiv \rho(\rho^3-3\eps\,\rho-\eps \la |c|),
\eeq
which gives for the metric
\beq\label{finmet}
g=\frac{|c|}{\rho}\left(\si_1^2+\ga^2\,\si_2^2+\frac 1{\ga^2}\,\si_3^2
+\frac 34\,\eps\,\frac{d\rho^2}{P(\rho)}\right),\qq\quad \ga^2\equiv e^{2|c|t}.
\eeq

\nin{\bf Remark:} Due to the symmetric role played by $\,(\si_2,\,\si_3)$, the coefficients of 
$\si_2^2$ and of $\si_3^2$ may be interchanged and this corresponds to the exchange 
$\,(c\leftrightarrow -c)$ or $\,\dst\Big(\ga\leftrightarrow\frac 1{\ga}\Big)$. This means that if 
the metric (\ref{finmet}) is Einstein, then 
\beq
g=\frac{|c|}{\rho}\left(\si_1^2+\frac 1{\ga^2}\,\si_2^2+\ga^2\,\si_3^2
+\frac 34\,\eps\,\frac{d\rho^2}{P(\rho)}\right),
\eeq
will be Einstein too. We will use this observation to get rid of the sign in relation (\ref{deft}) and to take $c>0$. 

Let us observe that the integration of relation (\ref{deft}) will require the use of elliptic 
functions. The corresponding reductions are given in the appendix; using these results we get the 
final form of the metrics, according to their signature. 

\section{Minkowskian signature}
In this case $P(\rho)=\rho(\rho^3+3\rho+\la c)$ has, no matter what the value of $c$ is, always 2 real and 2 complex conjugate roots (recall that we exclude $\la=0$). So we fix $c=1$ and, to express 
most conveniently the roots of $P$, we parametrize the Einstein constant according to
\[\la=2\,\sinh(\tht),\qq\qq \tht\in\,{\mb R}\backslash \{0\}.\]
We will use now the results from appendix B to give the explicit form of the metric.
\brm
\item \fbox{For $\la<0$ :} 

\nin In this case the roots are
\[a=-2\,{\rm sh}\,(\tht/3)\ >\ b=0,\qq a_1=\sqrt{3}\,{\rm ch}\,(\tht/3),\quad  
b_1={\rm sh}\,(\tht/3),\]
so we have
\[\left\{\barr{l}\dst 
A=\sqrt{3+12\,\sinh^2(\tht/3)},\\[4mm]\dst  B=\sqrt{3+4\,\sinh^2(\tht/3)},\earr\right.  
\qq\quad k^2=\frac{(A+B)^2-4\,\sinh^2(\tht/3)}{4AB}\]
and
\[{\rm sn}\,v_0=\sqrt{\frac{2B}{A+B-2\,\sinh(\tht/3)}}.\]
In formula (\ref{finmet}) we have to transform $d\rho$ into $dv$ to get eventually
\beq\label{finmet2}
g_M=\frac 1{\rho}\left(\si_1^2+\ga^2\,\si_2^2+\frac 1{\ga^2}\,\si_3^2
-\frac 3{AB}\,(dv)^2\right),\qq\quad v\in\,[0,v_0),
\eeq
where $\rho$ and $\ga^2$ are given respectively by
\beq
\rho=\frac{aB\,{\rm cn}^2\,v}{B\,{\rm cn}^2\,v-A\,{\rm sn}^2\,v\,{\rm dn}^2\,v},
\eeq
and by
\beq
\ga^2=\left(e^{-\xi v}\frac{H(v_0+v)\,\T_1(v_0+v)}{H(v_0-v)\,\T_1(v_0-v)}\right)^{\sqrt{3}},\qq 
\xi=2\left(\frac{\T'}{\T}(v_0)+\frac{H_1'}{H_1}(v_0)\right).
\eeq

\item \fbox{For $\la>0$ :} 

\nin In this case the roots are
\[a=0\ >\ b=-2\,{\rm sh}\,(\tht/3),\qq a_1=\sqrt{3}\,{\rm ch}\,(\tht/3),\quad  
b_1={\rm sh}\,(\tht/3),\]
so we have
\[\left\{\barr{l}\dst 
A=\sqrt{3+4\,\sinh^2(\tht/3)},\\[4mm]\dst  B=\sqrt{3+12\,\sinh^2(\tht/3)},\earr\right.  
\qq\quad k^2=\frac{(A+B)^2-4\,\sinh^2(\tht/3)}{4AB}.\]
The parameter $k^2$ remains unchanged while $A$ and $B$ are interchanged and $v_0$ becomes
\[{\rm sn}\,v_0=\sqrt{\frac{2B}{A+B+2\,\sinh(\tht/3)}}.\]
The metric is still given by (\ref{finmet2}), where now $\rho$ and $\ga^2$ are respectively 
\beq
\rho=\frac{|b|A\,{\rm sn}^2\,v\,{\rm dn}^2\,v}{B\,{\rm cn}^2\,v-A\,{\rm sn}^2\,v\,{\rm dn}^2\,v},
\eeq
and by
\beq
\ga^2=\left(e^{-\xi v}\frac{H(v_0+v)\,\T_1(v_0+v)}{H(v_0-v)\,\T_1(v_0-v)}\right)^{\sqrt{3}},\qq 
\xi=2\left(\frac{|b|}{AB}+\frac{\T'}{\T}(v_0)+\frac{H_1'}{H_1}(v_0)\right).\eeq
\erm

\section{Euclidean signature}
In this case $\,P(\rho)=\rho(\rho^3-3\rho-\la c)$.  It has two real roots for 
$\,\la c\in(-\nf,-2)\cup (+2,+\nf)$, four real roots for $\la c\in[-2,0)\cup(0,+2]$ and 
a double root for $\la c=\pm 2$. Since the parameter $c$ is free, we can collapse 
$\,(-\nf,0)\cup (0,+\nf)$ to two points by taking $c=2/|\la|$. In this case elliptic functions 
are no longer required, leading to simpler metrics.  

We have to discuss two cases:
\brm
\item \fbox{$\la<0$ :}

\nin We have $P(\rho)=\rho(\rho+2)(\rho-1)^2$ and
\[\frac{2c}{\sqrt{3}}\ dt=\frac{\rho\,d\rho}{|\rho-1|\sqrt{\rho(\rho+2)}}.\]
The change of variable $\dst\rho=\frac{2s^2}{3-s^2}$ simplifies to
\[2c\ dt=\frac{4s^2\,ds}{(1-s^2)(3-s^2)}.\]
We obtain
\beq
\ga^2\equiv e^{2ct}=\frac{1+s}{|1-s|}\left(\frac{\sqrt{3}-s}{\sqrt{3}+s}\right)^{\sqrt{3}},
\eeq
and the Einstein metric  
\beq
g_E=\frac{(3-s^2)}{|\la|\,s^2}\left(\si_1^2+\ga^2\,\si_2^2+\frac 1{\ga^2}\,\si_3^2
+\frac{ds^2}{(1-s^2)^2}\right).
\eeq
In fact we have two metrics: the first one for $\, s\in(0,1)$, and the second one for 
$\, s\in(1,\sqrt{3})$.

\item \fbox{$\la>0$ :}

\nin We have $P(\rho)=\rho(\rho-2)(\rho+1)^2$ and
\[\frac{2c}{\sqrt{3}}\ dt=\frac{\rho\,d\rho}{(\rho+1)\sqrt{\rho(\rho+2)}},\qq\quad \rho>2.\]
The change of variable $\dst\rho=\frac 2{1-s^2}$ simplifies to
\[\frac{2c}{\sqrt{3}}\ dt=-\frac{4\,ds}{(1-s^2)(3-s^2)},\qq s\in(-1,+1).\]
Deleting the sign we obtain
\beq
\ga^2\equiv e^{2c\,t}=\frac{\sqrt{3}-s}{\sqrt{3}+s}\left(\frac{1+s}{1-s}\right)^{\sqrt{3}}
\eeq
and the Einstein metric 
\beq
g_E=\frac{(1-s^2)}{\la}\left[\si_1^2+\ga^2\,\si_2^2+\frac 1{\ga^2}\,\si_3^2
+\frac{3\,ds^2}{(3-s^2)^2}\right]. 
\eeq

\erm

\section{Conclusion}
We have obtained some new Einstein metrics for Bianchi III and V. For this last case the 
complexity of the results remains reasonable since we end up simply with elliptic functions 
and not Painlev\'e transcendents. 

A very unusual ``bifurcation"' is observed: while in the minkowskian we need elliptic functions, in the euclidean one can dispense with them. This raises the following question: 
would it be possible, through clever changes, to get rid of the elliptic functions for all the 
Bianchi V Einstein metrics? Another question of interest is to what extent one could work out   
the more general Bianchi VI$_h$ and Bianchi VII$_h$ cases.

\vspace{1cm}
\centerline{\bf\Large Appendix}

\appendix

\section{De Sitter metric re-visited}
We have shown that the metric
\beq\label{dS1}
g=s^2\Big(\si_1^2+\si_2^2+\si_3^2\Big)-\frac{ds^2}{1+\frac{\dst\la\,s^2}{3}},
\eeq
is, for $\la>0$ de Sitter and for $\la<0$ anti-de Sitter. We will discuss only de Sitter. Taking 
${\rm sh}\,\tht=\sqrt{\frac{\la}{3}}\,s$ as a new variable the metric becomes
\beq\label{dS2}
g=\frac 3{\la}\Big({\rm sh}^2\,\tht(\si_1^2+\si_2^2+\si_3^2)-d\tht^2\Big),\qq\quad \la>0.
\eeq
This is quite a simple form fo de Sitter, which could be useful in other applications. So we will 
examine the isometries.

Let us first observe that the three dimensional metric
\[\si_1^2+\si_2^2+\si_3^2=dx^2+e^{2x}(dy^2+dz^2),\]
has 6 Killing vectors, shared by the metric (\ref{dS2}). It is made up with 2 sub-algebras:
\beq
{\cal A}_1=\Big\{P_1,\,P_2,\,M_3\Big\},\qq\qq {\cal A}_2=\Big\{Q_1,\,Q_2,\,L_3\Big\}.\eeq
The first one is $e(2)$ ($M_3$ is a rotation) 
\beq\barr{l}
P_1=\pt_y,\quad P_2=\pt_z,\quad M_3=-z\,\pt_y+y\,\pt_z,\\[5mm]
[M_3,P_1]=-P_2,\qq [M_3,P_2]=P_1,\qq [P_1,P_2]=0,\earr
\eeq
and the second one ($L_3$ mixes translation and dilatation)
\beq\left\{\barr{l}
Q_1=y\,\pt_x+\frac 12\Big(-y^2+z^2+e^{-2x}\Big)\pt_y-yz\,\pt_z,\\[4mm]  
Q_2=z\,\pt_x-yz\,\pt_y+\frac 12\Big(y^2-z^2+e^{-2x}\Big)\pt_z,\earr\right. \qq 
L_3=\pt_x-y\,\pt_y-z\,\pt_z, 
\eeq
with
\beq
[L_3,Q_1]=-Q_1,\qq [L_3,Q_2]=-Q_2,\qq [Q_1,Q_2]=0.
\eeq
These 2 sub-algebras close up according to
\beq\barr{ll}
[M_3,Q_1]=-Q_2,\qq & [M_3,Q_2]=Q_1,\\[4mm]
[L_3,P_1]=P_1,\qq & [L_3,P_2]=P_2, \\[4mm] 
[P_1,Q_1]=L_3,\qq &  [P_1,Q_2]=M_3,\\[4mm]
[P_2,Q_1]=-M_3,\qq & [P_2,Q_2]=L_3,\earr \qq \qq  [M_3,L_3]=0.
\eeq
We need 4 extra Killing vectors to get the 10 dimensional $o(4,1)$ Lie algebra for de Sitter metric. They are given by
\beq\barr{l}\dst 
C_1=e^x\left(\frac 1{{\rm th}\,\tht}\,\frac{\pt}{\pt x}-\,\frac{\pt}{\pt \tht}\right),\\[5mm]\dst 
C_2=ye^x\left(\frac 1{{\rm th}\,\tht}\,\frac{\pt}{\pt x}-\frac{\pt}{\pt \tht}\right)
+\frac{e^{-x}}{{\rm th}\,\tht}\,\frac{\pt}{\pt y},\quad  
C_3=ze^x\left(\frac 1{{\rm th}\,\tht}\,\frac{\pt}{\pt x}-\frac{\pt}{\pt \tht}\right)
+\frac{e^{-x}}{{\rm th}\,\tht}\,\frac{\pt}{\pt z},\\[5mm]\dst
C_4=\frac{(y^2+z^2)}{2}e^x\left(-\frac 1{{\rm th}\,\tht}\frac{\pt}{\pt x}+\frac{\pt}{\pt \tht}\right)
+\frac{e^{-x}}{{\rm th}\,\tht}\left(\frac 12\,\frac{\pt}{\pt x}-y\frac{\pt}{\pt y}-z\frac{\pt}{\pt z}
+\frac{{\rm th}\,\tht}{2}\frac{\pt}{\pt \tht}\right).\earr
\eeq
So despite the simple form of the metric, the isometries are quite awkward.

The remaining commutators (we give only the non-vanishing ones) are ordered as:
\beq\barr{lll}
[Q_1,C_1]=C_2\qq & [Q_2,C_1]=C_3\qq & [L_3,C_1]=C_1\\[4mm]
[P_1,C_2]=C_1\qq & [M_3,C_2]=-C_3\qq & [Q_1,C_2]=-C_4\\[4mm]
[P_2,C_3]=C_1\qq & [M_3,C_3]=C_2\qq & [Q_2,C_3]=-C_4\\[4mm]
[P_1,C_4]=-C_2\qq & [P_2,C_4]=-C_3\qq & [L_3,C_4]=-C_4\earr
\eeq
and 
\beq\barr{lll}
[C_1,C_2]=-P_1\qq & [C_1,C_3]=-P_2\qq & [C_2,C_3]=-M_3\\[4mm]
[C_2,C_4]=-Q_1\qq & [C_3,C_4]=-Q_2\qq & [C_1,C_4]=-L_3\earr
\eeq

\section{Elliptic functions: some tools}
There are plenty of books on elliptic function theory, but we used mainly the books by 
Byrd and Friedman \cite{bf} and by Whittaker and Watson \cite{ww}. We use Jacobi rather 
than Weierstrass notation for elliptic functions. Similarly we use earlier Jacobi notation 
for the theta functions which is best adapted to our purposes. They are related to 
the more symmetric notations used in \cite{ww} according to
\[H(v)=\tht_1(w),\quad H_1(v)=\tht_2(w),\quad  \T_1(v)=\tht_3(w),\quad \T(v)=\tht_4(w),
\qq w=\frac{\pi v}{2K}.\]

Let us start from the relation (\ref{deft})
\beq\label{deft2}
\frac{2dt}{\sqrt{3}}=\frac{\rho\,d\rho}{\sqrt{P(\rho)}}.
\eeq
If the quartic polynomial $P(\rho)$ has 2 real roots, and therefore two complex conjugate ones, we will write it
\[P(\rho)=(\rho-a)(\rho-b)[(\rho-b_1)^2+a_1^2],\qq a>b.\]
In this case, the positivity of $\rho$ and $P(\rho)$ requires $\rho\geq a$. One defines
\[A=\sqrt{(a-b_1)^2+a_1^2}\quad > \quad B=\sqrt{(b-b_1)^2+a_1^2},\qq k^2=\frac{(A+B)^2-(a-b)^2}{4AB}<1,\]
where $k^2$ will be the parameter of the elliptic functions involved. Let us define the change of variable
\beq\label{cofv}
{\rm sn}^2\,v=\frac{2B(\rho-a)}{D_+},\qq {\rm cn}^2\,v=\frac{D_-}{D_+},\qq  
{\rm dn}^2\,v=\frac{D_-}{2A(\rho-b)},\eeq
with
\beq
D_{\pm}=A(\rho-b)\pm B(\rho-a)+(a-b)\sqrt{(\rho-b_1)^2+a_1^2},\eeq
and the parameters
\[s_0\equiv {\rm sn}\,v_0=\sqrt{\frac{2B}{A+B+a-b}}<1,\quad 
s_1\equiv {\rm sn}\,v_1=\sqrt{\frac{2B}{A+B-a+b}}>1,\]
for which the reader can check that $\,v_1=K+iK'+v_0$. 

The change of variable (\ref{cofv}) transforms $\,\rho\in\,[a,+\nf)$ into $\,v\in\,[0,v_0)\subset [0,K_0)$.
 The inverse relation is \footnote{From now on we will use the simplified notations 
$s\equiv{\rm sn}\,(v,k^2),\ c\equiv{\rm cn}\,(v,k^2),\ d\equiv{\rm dn}\,(v,k^2)$ as well as 
$s_0={\rm sn}\,v_0,\ s_1={\rm sn}\,v_1$ etc... }
\beq\label{rho}
\rho=\frac{aB\,c^2-bA\,s^2 d^2}{B\,c^2-A\,s^2 d^2}.
\eeq
Using  
\[\barr{c}\dst 
\frac{\rho-a}{a-b}=\frac{A\,s^2 d^2}{B\,c^2-A\,s^2 d^2},\quad  
\frac{\rho-b}{a-b}=\frac{B\,c^2}{B\,c^2-A\,s^2 d^2},\\[5mm]\dst    
\sqrt{(\rho-b_1)^2+a_1^2}=AB\ \frac{d^2-c^2+c^2 d^2}{B\,c^2-A\,s^2 d^2},\earr\]
straightforward computations give
\[\frac{d\rho}{\sqrt{P(\rho)}}=\frac 2{\sqrt{AB}}\ dv.\]
It remains to give the explicit form of $\ga^2=e^{2t}$ as a function of $v$ by 
integrating (\ref{deft2}), which becomes now: 
\beq
\frac{2dt}{\sqrt{3}}=\frac 2{\sqrt{AB}}\frac{aB\,c^2-bA\,s^2\,d^2}{B\,c^2-A\,s^2\,d^2}\,dv.\eeq
The relation
\[\frac{c_0^2}{s^2-s_0^2}=-\frac{c_0}{2s_0d_0}\left(\frac{H'}{H}(v_0-v)+\frac{H'}{H}(v_0+v)
-2\frac{\T'}{\T}(v_0)\right),\]
and a similar one, obtained by the substitution $v_0\to v_1=K+iK'+v_0$:
\[\frac{c_1^2}{s^2-s_1^2}=\frac{c_0}{2s_0d_0}\left(\frac{\T_1'}{\T_1}(v_0-v)+\frac{\T_1'}{\T_1}(v_0+v)
-2\frac{H_1'}{H_1}(v_0)\right),\]
allow us to integrate up to
\beq
\ga^2\equiv e^{2t}=\left(e^{-\xi v}\frac{H(v_0+v)\,\T_1(v_0+v)}{H(v_0-v)\,\T_1(v_0-v)}\right)^{\sqrt{3}},
\qq \xi=2\left(-\frac b{\sqrt{AB}}+\frac{\T'}{\T}(v_0)+\frac{H_1'}{H_1}(v_0)\right).
\eeq
As the reader may notice, in \cite{bf}[p. 135] a different change of variables is given, which 
differs from ours. It is
\[{\rm cn}\, u=\frac{(A-B)\rho-bA+aB}{(A+B)\rho-bA-aB}.\]
As a consequence we get in the metric (\ref{finmet}) the term
\[-\frac 34\,\frac{d\rho^2}{P(\rho)}=-\frac 3{AB}\,\Big(\frac{du}{2}\Big)^2.\]
To avoid the $1/4$ factor we have used a duplication transformation to switch to our 
variable by $u=2v$. Notice that in the limit $\la\to\,0$ we have $3/AB\to\,1$.

\end{document}